# Supercontinuum generation in 2D graphene oxide film coated SiN waveguides

*David J. Moss[1]*

[1]Optical Sciences Centre, Swinburne University of Technology, Hawthorn, VIC 3122, Australia

## Abstract

Enhanced supercontinuum generation (SCG) is experimentally demonstrated in integrated silicon nitride ($Si_3N_4$) waveguides incorporating highly nonlinear graphene oxide (GO) in the form of two-dimensional (2D) films. On-chip integration of the 2D GO films with precise control of their thickness is realized by using a transfer-free and layer-by-layer coating method. The control of the film length and coating position is achieved via window opening in the upper silica cladding of the photonic integrated chips. Detailed SCG measurements are performed using the fabricated devices with different waveguide geometries and GO film thicknesses, and the results are compared with devices without GO. Significantly improved spectral broadening of ultrashort optical pulses with ultrahigh peaks powers exceeding 1000 W is observed for the hybrid devices, achieving up to 2.4 times improvement in the spectral bandwidth relative to devices without GO. Theoretical analyses for the influence of GO film thickness, coating length, coating position, and waveguide geometry are also provided by fitting the experimental results with theory, showing that there is still significant room for further improvement. This work opens up a promising new avenue towards improving the SCG performance of photonic integrated devices by incorporating functional 2D materials.

## 1. Introduction

Supercontinuum generation (SCG) describes the extreme spectral broadening of ultrashort optical pulses after propagation through a nonlinear medium, typically resulting from the interplay of a series of nonlinear optical processes [1-3]. It has enabled a novel generation of broadband optical sources that have wide applications in a variety of fields, such as metrology [4, 5], spectroscopy [6, 7], optical communications [8, 9], fluorescence microscopy [10, 11], optical coherence tomography [12, 13], and sensing [14, 15]. By using a single broadband source based on SCG to replace a collection of separate lasers for different wavelength bands, the size and complexity of systems can be greatly reduced [16, 17]. More importantly, SCG provides a powerful solution to access wavelengths where direct photon transitions cannot be efficiently excited via existing gain materials [18, 19]. The high coherence of the generated spectra across broad bandwidths can also be exploited for realizing self-referenced laser frequency combs [20, 21], coherent optical communications [8, 21], and precise frequency synthesis [22, 23].

Since the first discovery of SCG in bulk crystals in 1970 [24], SCG has been demonstrated based in many device platforms, which can be classified into two main categories. The first is optical fiber based platforms, such as silica fibers [3, 25] (including those incorporating specific dopants [26, 27]), fluoride fibers [28, 29], and $As_2S_3$ fibers [30, 31]. The second is photonic integrated platforms, mainly including silicon [16, 32], silicon nitride ($Si_3N_4$) [33, 34], silicon rich nitride [18, 35], chalcogenide glass [36, 37], high-index doped silica glass (Hydex) [38], aluminum gallium arsenide (AlGaAs) [39, 40], and silicon germanium alloys [41, 42]. Compared to off-chip optical fiber devices, photonic integrated platforms allow for the implementation of compact SCG devices on a chip scale, thus harvesting great dividends for integrated devices such as high stability and scalability, low power consumption, and large-scale manufacturing [43, 44]. Although silicon has been a dominant platform for linear photonic integrated devices [45, 46], its strong two photon absorption (TPA) in the near infrared telecommunications wavelength region and the associated parasitic free carrier absorption result in high nonlinear loss [43, 47], thus impeding the use of silicon photonic devices for efficient SCG when pumped in this wavelength range. Other integrated platforms have much lower TPA at near infrared wavelengths, although they still face the limitation of having a much smaller third-order optical nonlinearity than silicon [48, 49].

Recently, we have demonstrated enhanced nonlinear optical performance with on-chip integrated 2D graphene oxide (GO) films to overcome the intrinsic limitations arising from the material properties of existing photonic integrated platforms [50-56]. GO has shown many attractive advantages for implementing hybrid photonic integrated devices with high nonlinear optical performance, including a high third-order nonlinearity (about $10^4$ times that of silicon [51, 57]), relatively low linear loss compared to other 2D materials (over 100 times lower than that of graphene [52, 58]), a large optical bandgap (typically between 2.1 eV – 3.6 eV [59, 60]) that yields low TPA at near infrared wavelengths), facile fabrication processes with precise control [53, 61], and high compatibility with different integrated platforms [50, 62]. Based on these, we have demonstrated enhanced self-phase modulation (SPM) of optical pulses (with peak powers ≤ 160 W) in silicon and $Si_3N_4$ waveguides integrated with 2D GO films [54-56].

In this paper, we investigate the SCG in $Si_3N_4$ waveguides integrated with 2D layered GO films. We achieve on-chip integration of 2D GO films with precise control of their thickness, length, and placement. We perform detailed characterization of the SCG performance of the fabricated devices using ultrashort optical pulses with a maximum peak power of 1100 W. The experimental results show that the optical pulses propagate through the hybrid waveguides with significantly improved spectral broadening, achieving a maximum 15-dB bandwidth of ~217.5 nm, representing a 2.4 times improvement compared to devices without GO. By fitting the experimental results with theory, we provide an analysis of the influence of the GO film thickness, coating length and position, and waveguide geometry on the SCG performance, showing that further improvement can be achieved by optimizing the device structural parameters. These results verify the high nonlinear optical performance of $Si_3N_4$ waveguides integrated with 2D GO films for broadband nonlinear optical applications.

## 2. Device design and fabrication

**Figures 1(a)** and **(b)** shows schematics of GO's atomic structure and bandgap. GO has a lattice-like nanostructure of hexagonal carbon rings which contains various oxygen-containing functional groups (OCFGs) such as epoxide, hydroxyl, and carboxylic, either in the basal plane or on the sheet edges [59, 63]. Due to its heterogeneous atomic structure, GO exhibits a series of distinctive material properties. Compared with graphene, GO provides a highly flexible material platform by manipulating the OCFGs to engineer its bandgap and hence properties, which has underpinned a variety of linear and nonlinear optical applications [50, 51]. In contrast to graphene that has a zero bandgap [64], GO has an opened bandgap stemming from the isolated $sp^2$ domain within the $sp^3$ C–O matrix. The bandgap of GO typically is between 2.1 eV – 3.6 eV [59, 65], which yields both low linear light absorption and a low nonlinear TPA in the near infrared wavelength region.

**Figure 1(c)** shows the schematic of a $Si_3N_4$ waveguide integrated with a monolayer GO film. In this work, we chose the $Si_3N_4$ platform since it has shown a good compatibility for hybrid integration of 2D GO films [52, 56, 66]. Compared to silicon that has a relatively small bandgap of ~1.1 eV [43, 67], $Si_3N_4$ has a large bandgap of ~5.0 eV [48, 68] that yields low TPA in the near infrared region. We fabricated $Si_3N_4$ waveguides via CMOS-compatible, annealing-free, and crack-free processes as reported previously [52, 69, 70]. A 2.3-µm thick silica layer was deposited on the fabricated $Si_3N_4$ waveguides as an upper cladding. Next, lithography and dry etching processes were employed to open a window in the silica cladding down to the top surface of the $Si_3N_4$ waveguides, allowing for subsequent GO film coating. The coating of 2D GO films was achieved by using a solution-based and transfer-free film coating method [61], where self-assembled GO films were constructed layer by layer by repeating four procedures needed for coating monolayer GO films. This method enables the fabrication of large-area GO films with well controlled thickness on a nanometer scale. It has been employed for the on-chip integration of GO films to realize functional devices [50, 58, 71], showing high repeatability, scalability, and compatibility with different integrated platforms.

**Figure 1(d)** shows a microscope image around the window opening area of the fabricated $Si_3N_4$ chip coated with a monolayer GO film. The coated GO film had a thickness of ~2 nm and showed good morphology and high transmittance. Since our GO film coating method is based on self-assembly via electrostatic attachment, it enables conformal coating of the GO film in the window opening area without any observable wrinkling and stretching, showing advantages as compared with film transfer methods widely employed for coating other 2D materials such as graphene and transition metal dichalcogenides (TMDCs) [62]. The opened window allows for the control of the coating length and placement of the GO films, which is important for optimizing the performance of the hybrid waveguides for different applications [50, 52, 58].

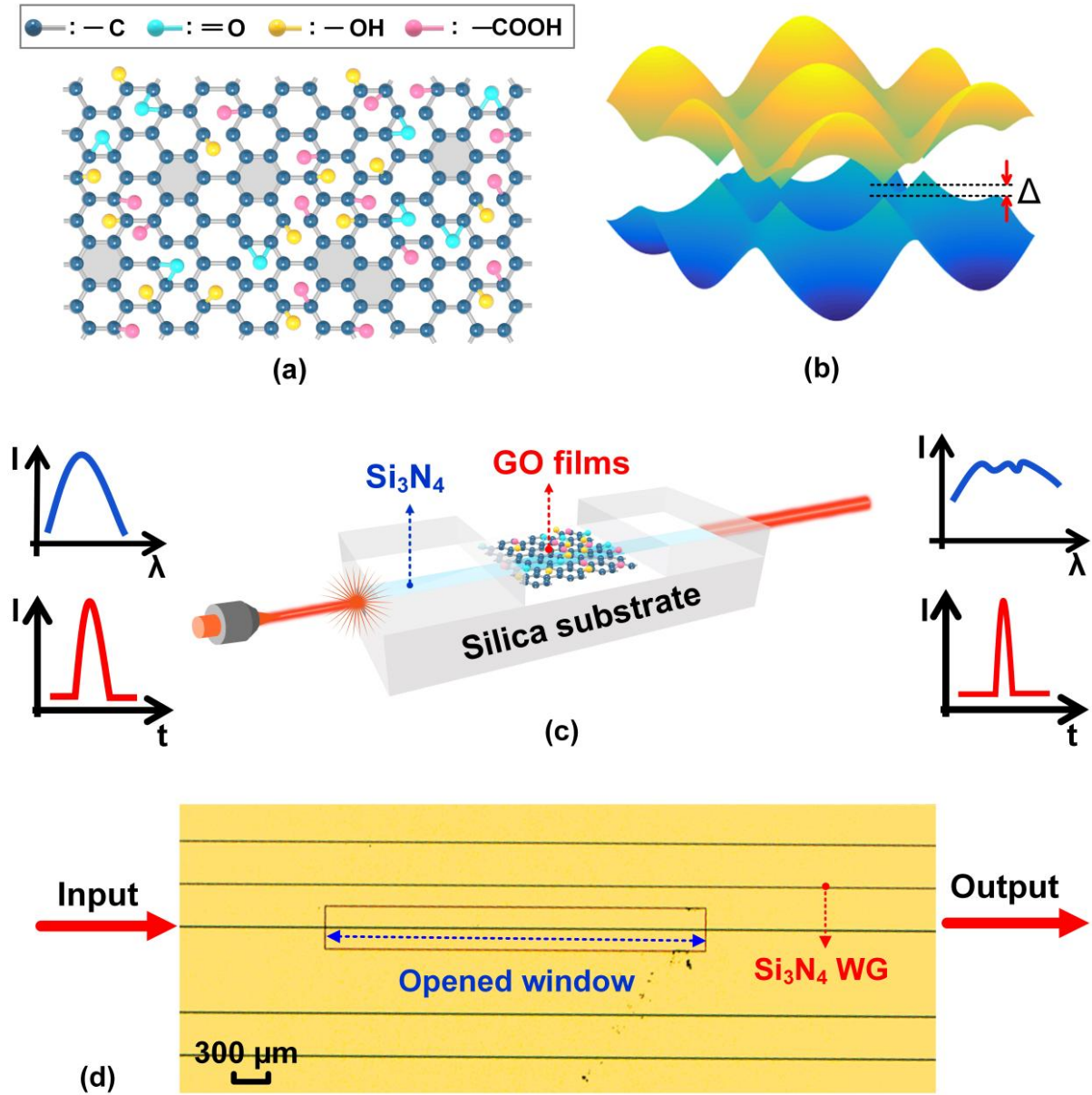

**Figure 1**. (a) Schematic of GO's atomic structure. (b) Schematic of GO's bandgap. (c) Schematic of a $Si_3N_4$ waveguide integrated with a monolayer GO film. (d) Microscope image of the fabricated $Si_3N_4$ integrated chip coated with a monolayer GO film. WG: waveguide.

**Figure 2(a-i)** shows a schematic cross section of the hybrid waveguide with a monolayer GO film. The corresponding transverse electric (TE) mode profile is shown in **Figure 2(a-ii)**. The $Si_3N_4$ waveguide had a width of $W$ = 1.60 μm, and a height of $H$ = 0.72 μm. The light-matter interaction between the highly nonlinear GO film and the waveguide evanescent field boosts the nonlinear optical response of the hybrid waveguide, forming the basis for the enhancement of the SCG performance. We chose TE-polarization for our SCG measurements since it supports in-plane interaction between the waveguide evanescent field and the GO film, which is much stronger compared to the out-of-plane interaction in light of the significant optical anisotropy in 2D materials [58, 72]. **Figure 2(b)** shows the refractive indices of GO, $Si_3N_4$, and silica ($SiO_2$) over a broad wavelength range of 1200 nm – 2000 nm measured by an ellipsometer. GO has a refractive index that is higher than $SiO_2$, but slightly lower than $Si_3N_4$, resulting in a strong mode overlap with the GO film in the hybrid waveguide.

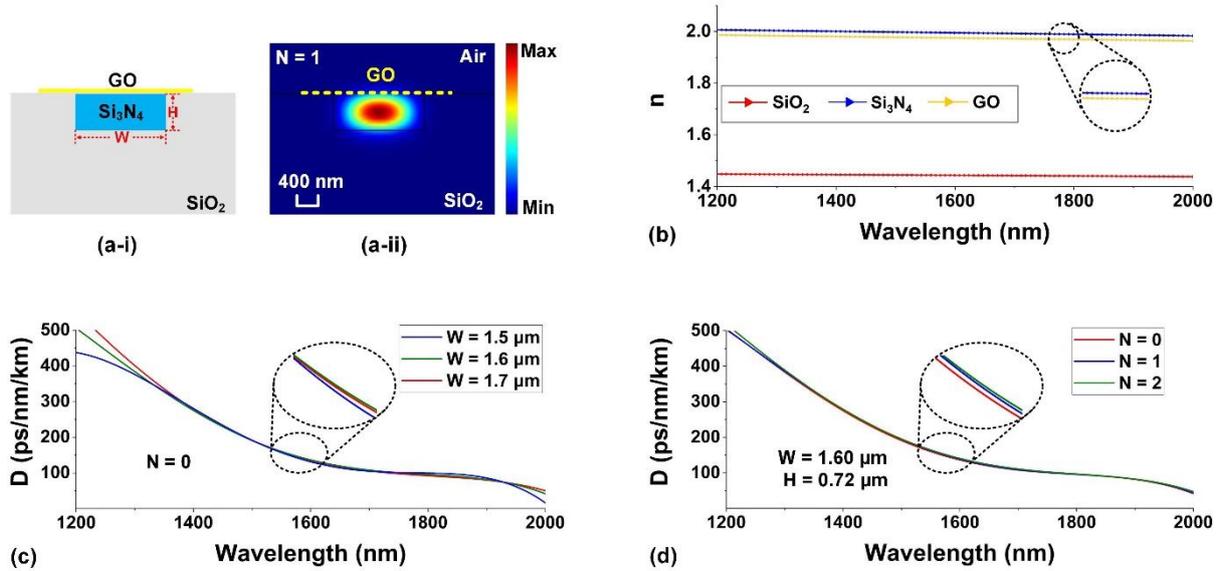

**Figure 2**. (a) Schematic cross section and corresponding TE mode profile of a $Si_3N_4$ waveguide integrated with a monolayer GO film. (b) Comparison of refractive indices $n$ of GO, $Si_3N_4$, and $SiO_2$ in the wavelength range of 1200 nm – 2000 nm. (c) Comparison of dispersion ($D$) of the uncoated $Si_3N_4$ waveguides with different widths ($W$ = 1.50, 1.60, and 1.70 μm) at a fixed height ($H$ = 0.72 μm). (d) Comparison of dispersion ($D$) of the uncoated waveguide ($N$ = 0) and hybrid waveguides with 1 and 2 layers of GO ($N$ = 1, 2) when $W$ = 1.60 μm and $H$ = 0.72 μm.

**Figure 2(c)** shows the dispersion ($D$) of the uncoated $Si_3N_4$ waveguides with different widths ($W$) at a fixed height ($H$) of 0.72 μm, which was simulated via commercial FDTD mode solution software based on the refractive indices in **Figure 2(b)**. We chose three different values of $W$ (i.e., 1.50 μm, 1.60 μm, and 1.70 μm) that are consistent with those of our fabricated devices. As can be seen, there is anomalous dispersion for all the three $Si_3N_4$ waveguides, which is critical for reducing the phase mismatch, thus being beneficial for the SCG. The dispersion can be engineered by changing the waveguide geometry, with slightly enhanced anomalous dispersion being achieved as $W$ increases from 1.50 μm to 1.70 μm. **Figure 2(d)** compares the dispersion of the uncoated waveguide and the hybrid waveguides with 1 and 2 layers of GO. For comparison, all of them have the same geometry ($W$ = 1.60 μm, $H$ = 0.72 μm) for the $Si_3N_4$ waveguide. After incorporating the GO films, the hybrid waveguides show enhanced anomalous dispersion compared to waveguides without GO. As the GO layer number $N$ increased, the anomalous dispersion of the hybrid waveguide was further enhanced.

## 3. Loss measurements

The experimental setup to characterize the linear and nonlinear loss of the fabricated devices is shown in **Figure 3(a)**. Coupling light to the device under test (DUT) was achieved with lensed fibers butt coupled to inverse-taper couplers at both ends of the waveguides, with a fiber-to-chip coupling loss of ~3 dB / facet. A polarization controller (PC) was employed to adjust the polarization of the input light to TE-polarized. Optical power meters (OPM 1 and OPM 2) were employed to measure the average power before and after the DUT, respectively. The input power was adjusted with a broadband variable optical attenuator (VOA), and an optical isolator was inserted to prevent damage to the laser.

The linear loss was measured with continuous-wave (CW) light while the nonlinear loss was measured with a fiber pulsed laser (FPL) with optical pulsewidth ~1.9 ps and repetition rate ~60 MHz, amplified by an erbium doped fiber amplifier (EDFA). **Figure 3(b)** shows the pulse duration ($\tau$) of the amplified pulses versus EDFA pump current ($I$) showing a minimum $\tau$ of ~0.2 ps at $I$ = ~160 mA at an average amplified output power of ~13.2 mW, corresponding to a peak power of ~1100 W, for an input pulses with ~0.3 mW average and ~3 W peak power. **Figure 3(c)** shows the optical spectrum of the optical pulses before and after amplification ($I$ = ~160 mA), showing 15-dB bandwidths of ~6 nm and ~73 nm, respectively.

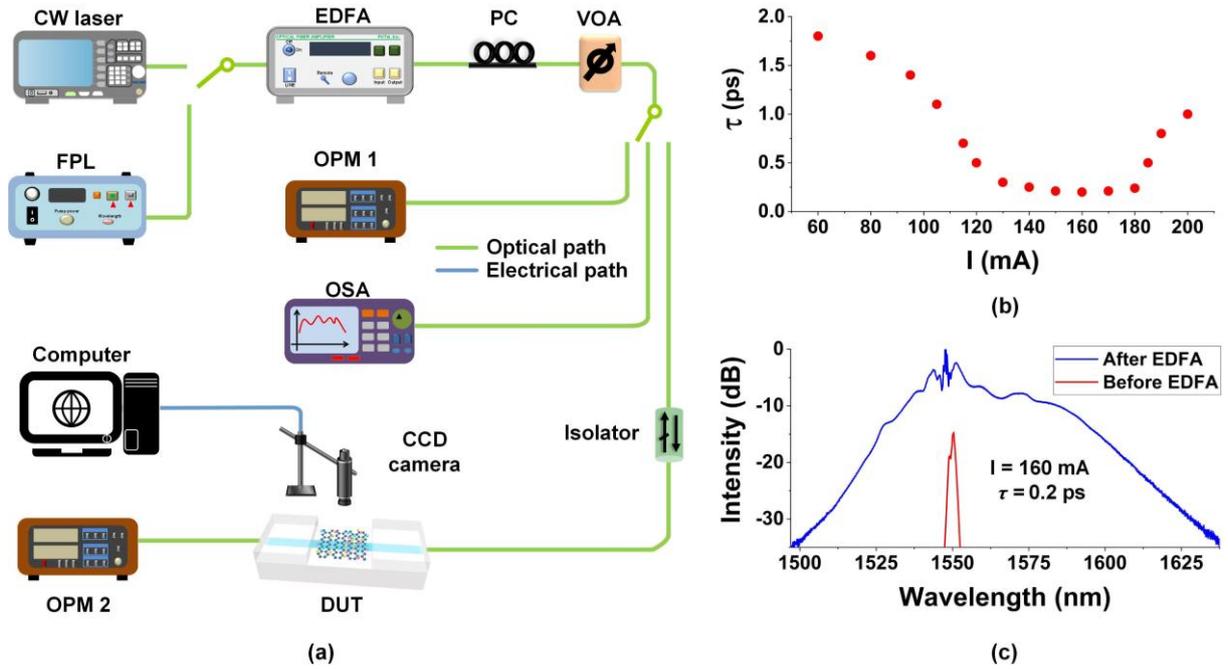

**Figure 3**. (a) Experimental setup for measuring linear and nonlinear loss of the GO-coated $Si_3N_4$ waveguides. CW laser: continuous-wave laser. FPL: fiber pulsed laser. PC: polarization controller. EDFA: Erbium doped fiber amplifier. VOA: variable optical attenuator. OPM: optical power meter. DUT: device under test. CCD: charged-coupled device. OSA: optical spectrum analyzer. (b) Pulse duration ($\tau$) of optical pulses amplified by the EDFA versus the EDFA's pump current ($I$). (c) Optical spectrum of the amplified optical pulses at the pump current of ~160 mA. The optical spectrum of the signal before amplification is also shown for comparison.

**Figure 4(a)** shows the insertion loss of the $Si_3N_4$ waveguides (20 mm long, $W$ = 1.60 μm and $H$ = 0.72 μm) for the uncoated ($N$ = 0) and hybrid waveguides with 1 and 2 layers of GO ($N$ = 1, 2). The hybrid waveguides with opened windows $L_c$ = 1.4 mm long were located at $L_0$ = 0.7 mm from the input port. The coupled input power of the CW light was ~0 dBm. The flat spectral response indicates the absence of wavelength-dependent material absorption or coupling loss for all three devices. From **Figure 4(a)** we extract the excess propagation loss of the GO films of ~3.0 dB/cm and ~6.1 dB/cm for the devices with 1 and 2 layers of GO, respectively. Compared to $Si_3N_4$ waveguides with graphene [70, 73], the loss induced by GO is about 2 orders of magnitude lower, mainly due to its large bandgap that results in low light absorption at infrared wavelengths. This is an important advantage of GO for SCG where high light intensity is needed.

**Figure 4(b)** shows the insertion loss of the three devices measured as a function of the input CW power, at a fixed wavelength of 1550 nm. For the hybrid waveguides, the insertion loss did not show any variation below 16 mW, indicating that the power-dependent loss induced by the photo-thermal changes in the GO films was negligible in this power range. This is consistent with previous results where the typical average power induced photo-thermal changes only occurred above 40 mW [50, 52, 74].

**Figure 4(c-i)** shows the insertion loss of the three devices with the amplified optical pulses at $I = \sim160$ mA. **Figure 4(c-ii)** shows the excess insertion loss ($\Delta EIL$) for the hybrid waveguides extracted from **Figure 4(c-i)** after excluding the insertion loss of the uncoated $Si_3N_4$ waveguide. The average power of the optical pulses was adjusted from 7.2 mW to 13.2 mW by the VOA, corresponding to peak powers of 600 W – 1100 W. Since photo-thermal changes are sensitive to the average power in the GO films [50, 52] which was < 16 mW, they were negligible. The nonlinear insertion loss of the hybrid waveguides decreased with increasing pulse peak power, with the 2 layer device showing a more significant decrease than the single layer device. In contrast, the insertion loss of the uncoated $Si_3N_4$ waveguide remained constant. These results indicate that the hybrid waveguides experienced saturable absorption (SA) in the GO films, similar to observations for waveguides incorporating graphene [70, 75].

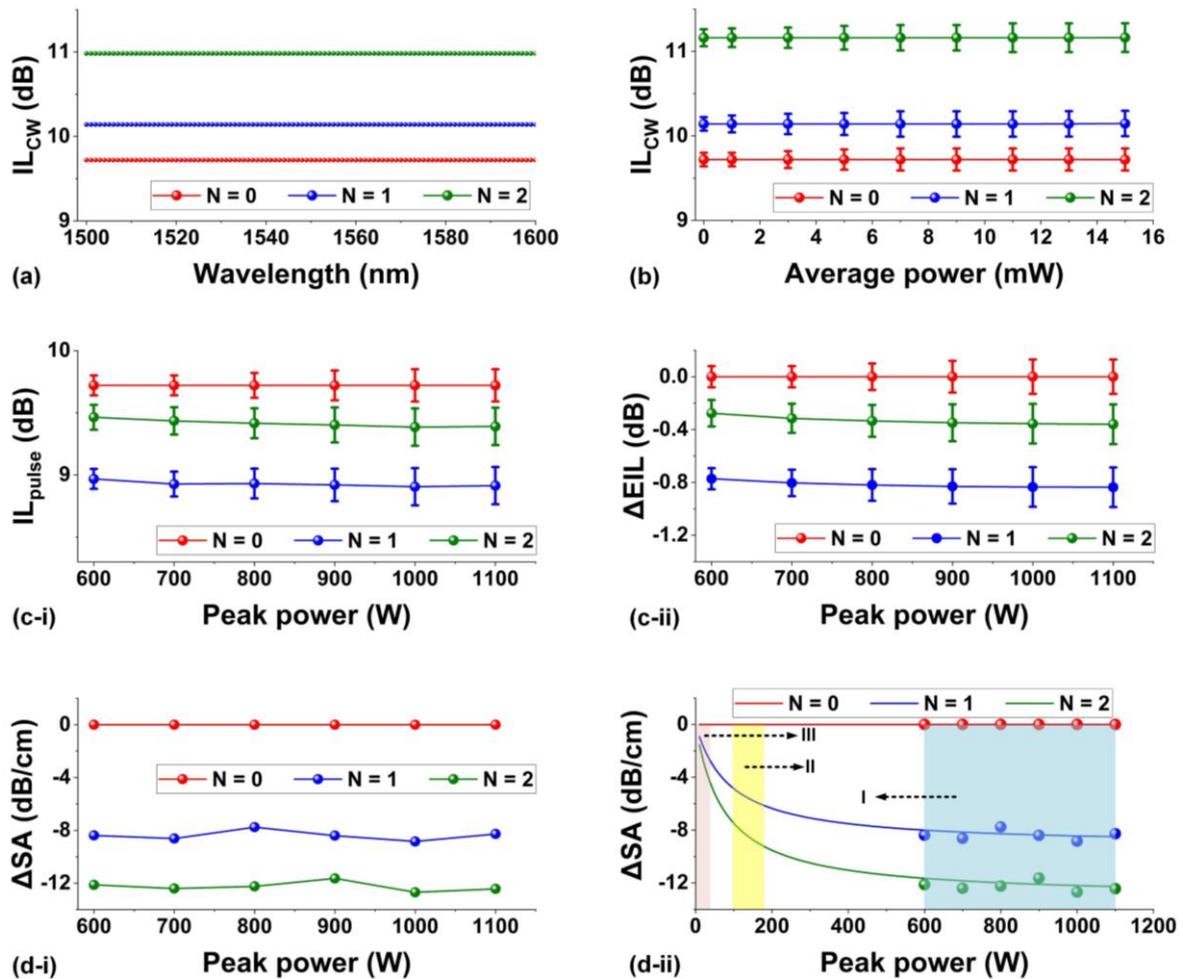

**Figure 4**. (a) Measured insertion loss ($IL_{CW}$) of the fabricated devices versus wavelength of the continuous-wave

(CW) light. The input CW power is 0 dBm. (b) Measured insertion loss ($IL_{CW}$) of the fabricated devices versus input CW power. The input CW wavelength is 1550 nm. (c-i) Measured insertion loss ($IL_{pulse}$) of the fabricated devices versus peak power of input optical pulses. (c-ii) Excess insertion loss ($\Delta EIL$) extracted from (c-i) after excluding the insertion loss of the uncoated $Si_3N_4$ waveguide. (d-i) Excess propagation loss induced by the saturable absorption in GO ($\Delta SA$) versus peak power of input optical pulses. (d-ii) $\Delta SA$ as a function of peak power of input optical pulses obtained by fitting the results in (d-i) with theory. I–III highlight the peak power ranges of the optical pulses used in this work and for SPM experiments in Refs. [54, 56], respectively. In (a) – (d), the curves for $N = 0, 1, 2$ correspond to the results for the uncoated $Si_3N_4$ waveguides, and the hybrid waveguides with 1 and 2 layers of GO, respectively.

**Figure 4(d-i)** shows the SA-induced excess propagation loss ($\Delta SA$, after excluding the linear propagation loss) versus the peak power of the input optical pulses, which was extracted from the results in **Figures 4(b)** and **(c)**. The negative values of $\Delta SA$ reflect the decrease in loss for an increased peak power in the SA process, which is beneficial for improving the nonlinear spectral broadening in the SCG process. **Figure 4(d-ii)** shows $\Delta SA$ for the hybrid waveguides with 1 and 2 layers of GO as a function of the input peak power, which was obtained by fitting the results in **Figure 4(d-i)** with the SA theory as follows [76, 77]:

$$\alpha_{SA} = \alpha_{sat} /(1 + \frac{|A|^2}{I_s}) \qquad (1)$$

where $\alpha_{SA}$ is the loss induced by GO, $\alpha_{sat}$ is the SA coefficient, $I_s$ is the saturation intensity, and $A$ is the field amplitude of the optical pulses. In **Figure 4(d-ii)**, the loss decrease becomes saturated at high peak powers > 400 W, showing a trend that is consistent with the inverse proportional function in **Eq. (1)**. We also labelled the peak power range of the amplified optical pulses (*I*) in this work, together with the those of the optical pulses (*II, III*) used in previous SPM experiments [54, 56]. As can be seen, the peak power of the amplified optical pulses here far exceeds those of the optical pulses employed for the SPM measurements [78, 79].

## 4. SCG measurements

We used the experimental setup in **Figure 5** to measure the SCG in the devices. The amplified optical pulses in **Figure 3(c)** were injected into the devices, and the spectral broadening of the output signal was recorded by an optical spectrum analyzer. To enhance the SCG the input pulses are normally launched around the zero-dispersion wavelengths or in the wavelength regime with anomalous dispersion [3, 41]. The former can reduce the temporal broadening of the optical pulses to maintain the high peak power for nonlinear optical interactions [33]. The latter, on the other hand, facilitates the propagation of solitons whose dynamics can be engineered to further enhance the spectral broadening [80]. In our SCG experiments, the center wavelength of the spectrally broad optical pulses was near 1550 nm. According to the results in **Figures 2(c)** and **(d)**, this is in the anomalous dispersion regime for both the bare and hybrid waveguides.

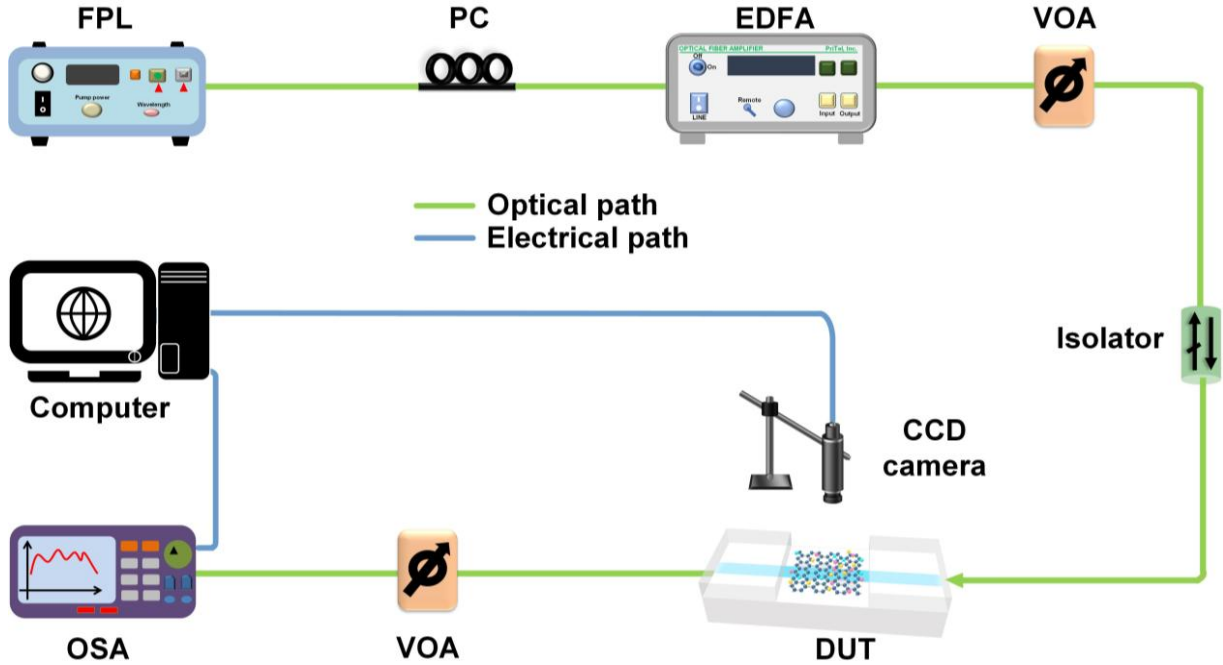

**Figure 5**. Experimental setup for measuring SCG in the GO-coated $Si_3N_4$ waveguides. FPL: fiber pulsed laser. PC: polarization controller. EDFA: Erbium doped fiber amplifier. VOA: variable optical attenuator. DUT: device under test. CCD: charged-coupled device. OSA: optical spectrum analyzer.

**Figure 6(a)** shows the normalized spectra of the optical pulses after propagation through the uncoated and GO-coated $Si_3N_4$ waveguides. The spectrum of the input optical pulses (peak power ~1100 W) is also shown for comparison. The output spectra from all devices show spectral broadening, with the hybrid waveguides showing more significant broadening compared to the uncoated waveguide, with the 2 layer device showing greater broadening than the single layer device. We would expect to see even greater spectral broadening for devices with more layers, similar to what we observed in our FWM experiments [52] since devices with thicker films have a significantly increased Kerr nonlinearity. This is accompanied by an increase in loss, however, and so finding a balance (in layer thickness) would optimize the performance.

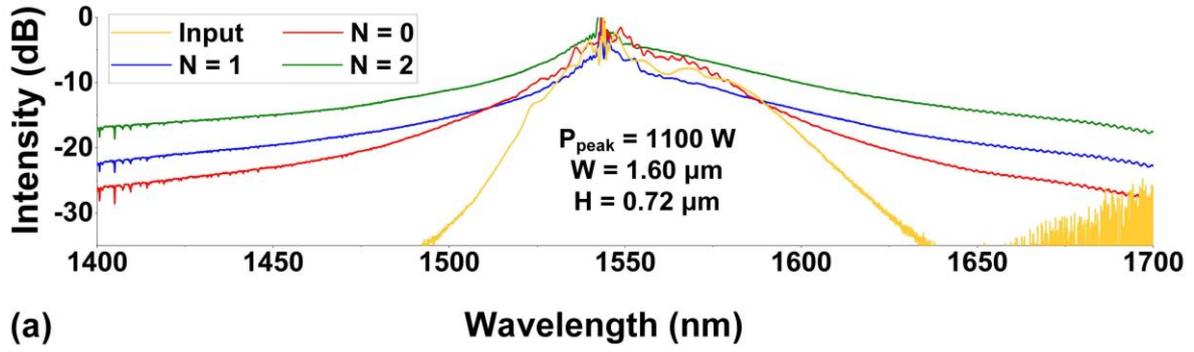

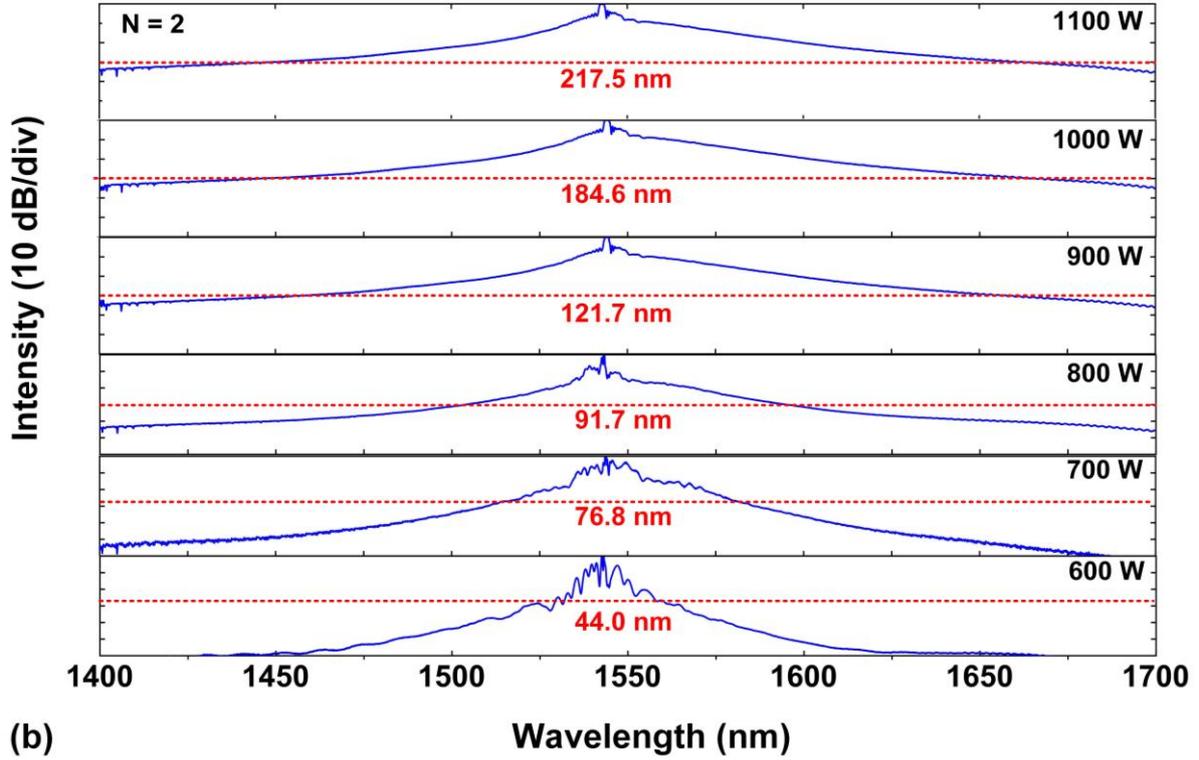

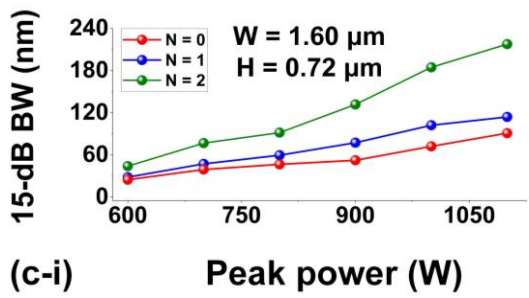 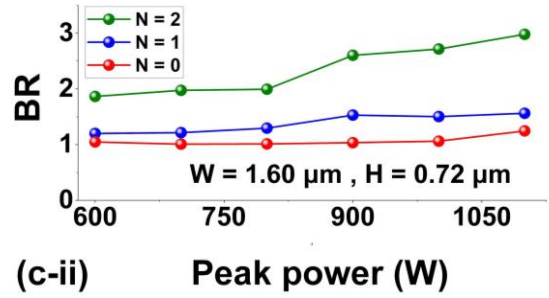

**Figure 6**. (a) Normalized spectra of optical pulses before and after propagation through uncoated ($N = 0$) and hybrid waveguides with 1 and 2 layers of GO ($N = 1, 2$) at an input peak power of ~1100 W. (b) Measured output spectra from the hybrid waveguides with 2 layers of GO at different input peak powers. The -15-dB bandwidth (BW) in each spectrum is also labelled. (c) -15-dB BWs and broadening ratios (BRs) of the output spectra from the uncoated and hybrid waveguides versus peak power of input optical pulses. In (a) – (c), all the devices have the same geometry ($W = 1.60$ μm, $H = 0.72$ μm) for the $Si_3N_4$ waveguides.

Given the range limit of our optical spectrum analyzer ($\lambda \leq 1700$ nm), we define the spectral bandwidth (BW) of the SCG at the -15-dB level. We also define a broadening ratio (BR) as:

$$BR = \frac{BW_{out}}{BW_{in}} \qquad (2)$$

where $BW_{in}$ and $BW_{out}$ are the -15-dB BWs of the input and output signals, respectively.

**Figure 6(b)** shows the output spectra from the hybrid waveguide with 2 layers of GO for input optical pulses with different peak powers. We varied the input peak power from 600 W to 1100 W, as in **Figure 4(c)**. As expected for the nonlinear SCG process, the spectral broadening of the output increases with increasing input peak power. **Figures 6(c-i)** and **6(c-ii)** shows the BWs and BRs of the output spectra from the uncoated and hybrid waveguides versus input peak power. Both are higher for the hybrid waveguide with 1 layer of GO than for the uncoated waveguide, but lower than the device with 2 layers of GO, in agreement with **Figure 6(a)**. Both the BW and the BR increase with the input peak power, which is consistent with **Figure 6(b)**. A maximum BW of ~217.5 nm was achieved for the device with 2 layers of GO, which is ~3.0 times that of the input optical pulse and ~2.4 times that of the output from the uncoated waveguide.

We also measured the SCG performance for devices with different waveguide geometries. **Figure 7(a)** compares the output spectra from three uncoated $Si_3N_4$ waveguides with the same height of $H$ = 0.72 μm but different widths of $W$ = 1.50 μm, 1.60 μm, and 1.70 μm. The waveguide geometries were the same as in **Figure 2(c)**, and the input peak power for all devices was ~1100 W. Spectral broadening was greater for the wider waveguides, in agreement with the simulation results in **Figure 2(c)**, and mainly arises from a stronger mode confinement within the waveguide core and the increased anomalous dispersion as $W$ increases.

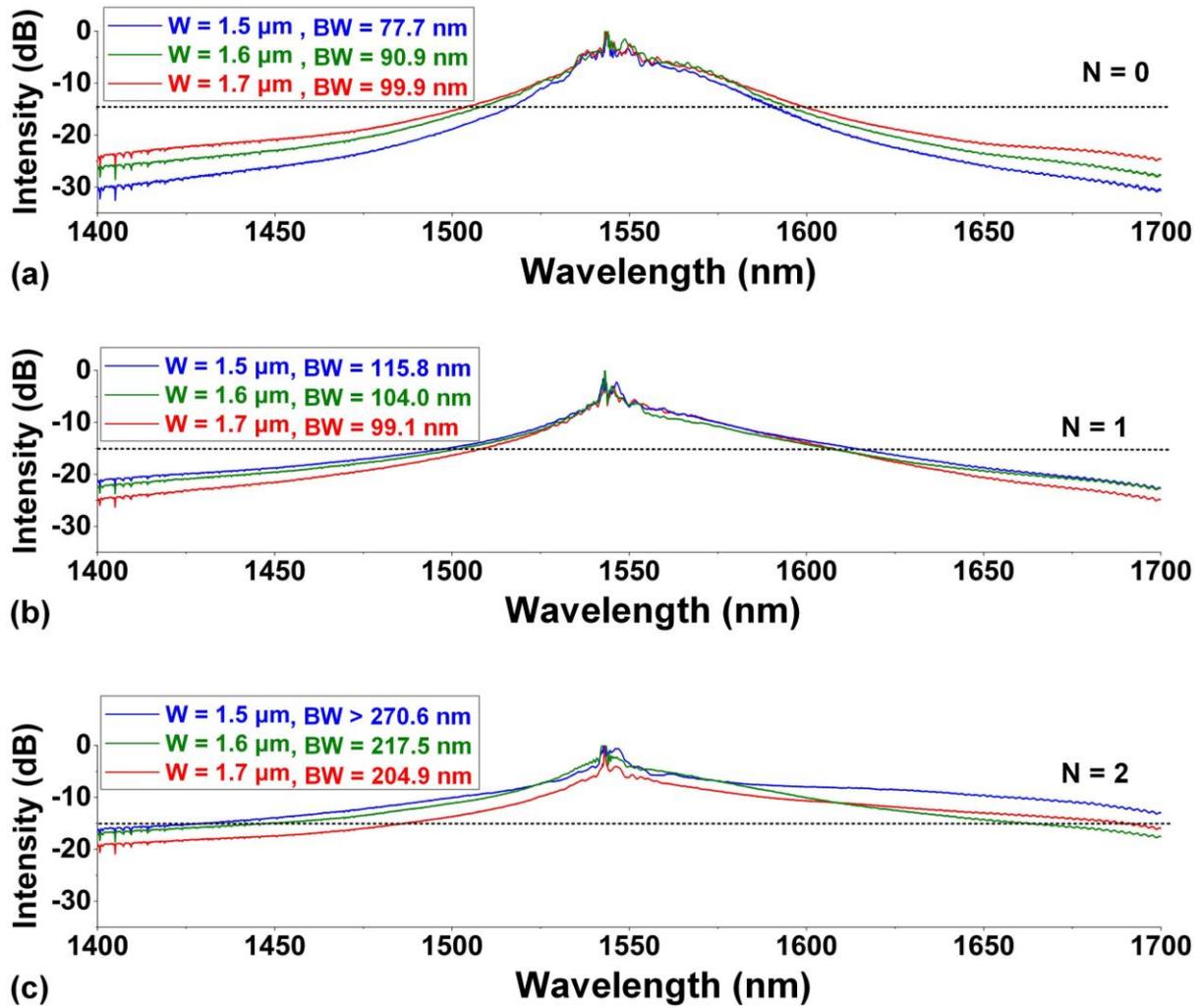

**Figure 7**. (a) Normalized spectra of optical pulses after propagation through uncoated $Si_3N_4$ waveguides with different widths of $W$ = 1.5, 1.6, and 1.7 μm. (b) − (c) Normalized spectra of optical pulses after propagation through hybrid waveguides with 1 and 2 layers of GO ($N$ = 1, 2) coating on the devices in (a). In (a) – (c), the waveguide height is $H$ = 0.72 μm and the peak power of the input optical pulses is ~1100 W. In each spectrum, the 15-dB bandwidth (BW) is also labelled. In (b) and (c), the GO film length and coating position are $L_c$ = 1.4 mm and $L_0$ = 0.7 mm, respectively.

**Figures 7(b)** and **(c)** show the results for the hybrid waveguides with 1 and 2 layers of GO, respectively. For the hybrid waveguides, the waveguide geometries were the same as **Figure 7(a)**, and the GO coating length and position were the same as **Figure 6**. As can be seen, the devices with 2 layers of GO show greater spectral broadening than comparable devices with 1 layer of GO. The narrower hybrid waveguides showed greater spectral broadening - a trend opposite to that in **Figure 7(a)**. This is mainly due to the increase of GO mode overlap as $W$ decreases. The spectral broadening induced by SCG is a combined result from several factors, including the optical nonlinearity, dispersion, and loss of the waveguides. Although the uncoated $Si_3N_4$ waveguide with a smaller $W$ had weaker SCG, the improvement in the optical nonlinearity of the hybrid waveguides arising from the increased mode overlap with the highly nonlinear GO films had a greater influence on the spectral broadening. These results further confirm the high third-order optical nonlinearity of the GO films.

**5. Theoretical analysis and discussion**

To analyze the experimental results in **Section 4**, we simulated the SCG-induced spectral broadening of optical pulses by numerically solving the following generalized nonlinear Schrödinger equation [3, 38]:

$$\frac{\partial A}{\partial z} = -\frac{1}{2}\alpha A + \sum i^{m+1}\frac{\beta_m}{m!}\frac{\partial^m A}{\partial t^m} + i\gamma(1 + \frac{i}{\omega_0}\frac{\partial}{\partial T})A \int R(T')|A(z,T-T')|^2 dT' \quad (3)$$

where $A(z, t)$ is the slowly varying field envelop of the optical pulses along the propagation direction $z$, $\alpha$ is the loss factor including both the linear and nonlinear loss in **Figure 4**, $\beta_m$ is is the $m$-th order dispersion coefficient of the waveguide propagation constant, $\gamma$ is the waveguide nonlinear parameter, $R(T)$ the Raman response function, and $T$ the time frame moving with the pulse group velocity ($T = t - z/v_g$). In our simulation, the GO-coated $Si_3N_4$ waveguides were divided into uncoated (with silica cladding) and hybrid segments (with GO films and air cladding). Numerically calculating **Eq. (3)** was performed for each segment, and the output from the previous segment was set as the input for the subsequent one.

SCG is a combined result of several different nonlinear optical processes [32, 81]. In contrast to silica optical fibers that have relatively strong stimulated Raman scattering [3, 82], the influence of the Raman-induced soliton fission and self-frequency shift on the SCG from $Si_3N_4$ waveguides is negligible in the near infrared region [34, 83]. For the hybrid waveguides incorporating 2D GO films, their SCG performance is also affected by the new features brought by the GO films, such as improved third-order optical nonlinearity, increased linear loss, and reduced loss arising from the SA.

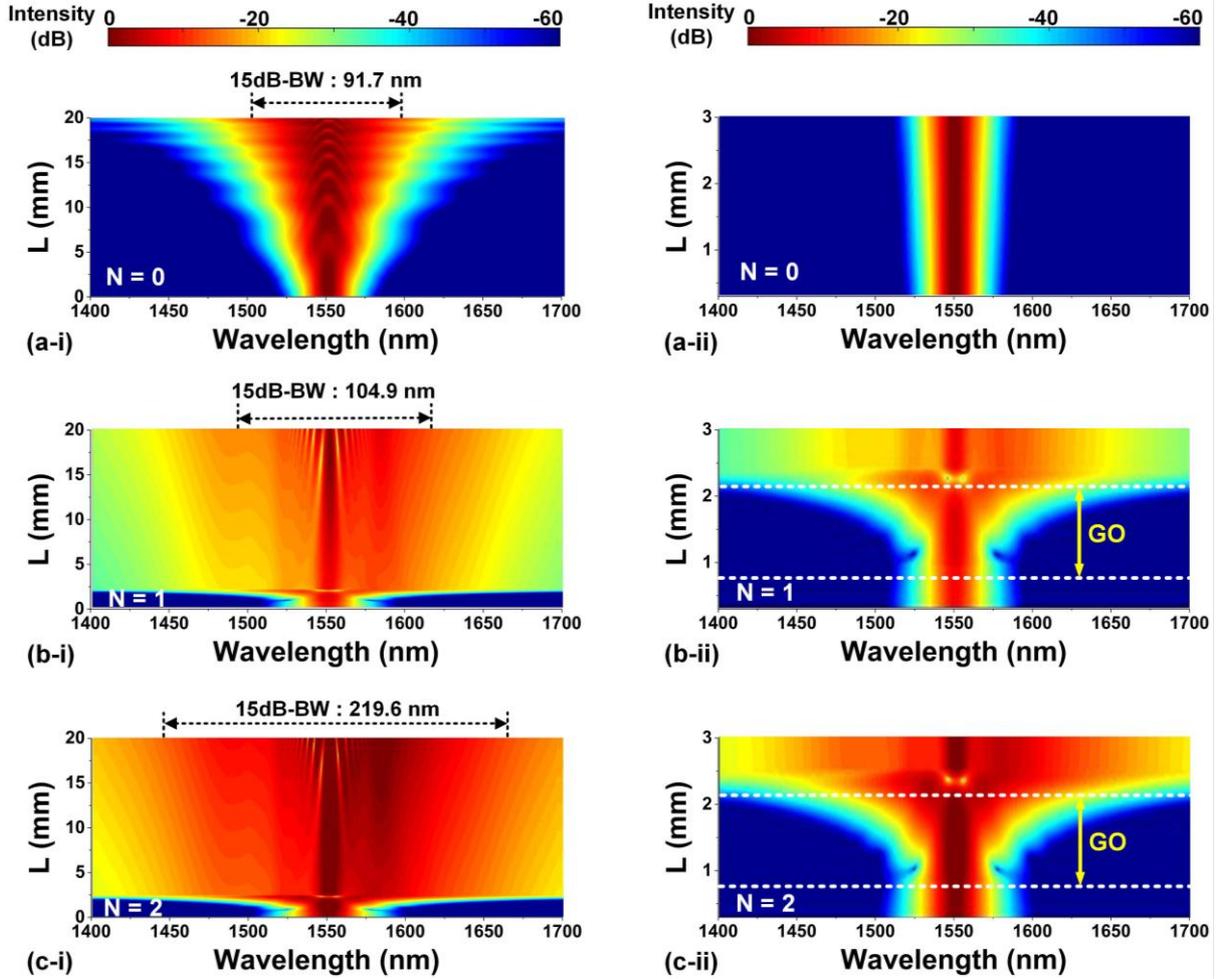

**Figure 8**. (a) Simulated spectrum evolution of the optical pulses propagating along the uncoated waveguide ($N = 0$). (b) – (c) Simulated spectrum evolution of the optical pulses propagating along the hybrid waveguides with 1 and 2 layers of GO ($N = 1, 2$), respectively. In (a) – (c), (i) shows the spectrum evolution along the entire waveguide (from 0 mm to 20 mm) and (ii) shows the zoom-in view around the window opening area (from 0.7 mm to 2.1 mm). The simulated spectrum evolution was obtained by fitting the results in Figure 6(a) at an input peak power of ~1100 W, and all the devices have the same geometry ($W = 1.60$ μm, $H = 0.72$ μm) for the $Si_3N_4$ waveguides. The 15-dB bandwidth (BW) is labelled in (i) for each figure.

**Figures 8(a-i) – (c-i)** shows the simulated spectral evolution of the optical pulses propagating along the uncoated and hybrid waveguides, obtained by fitting the experimental results in **Figure 6(a)**. **Figures 8(a-ii) – (c-ii)** show the corresponding zoom-in views around the window opening area (from 0.7 mm to 2.1 mm). The fit $γ$ of the uncoated $Si_3N_4$ waveguide ($W = 1.60$ μm and $H = 0.72$ μm) was ~1.51 $W^{-1}m^{-1}$, consistent with previously reported values for similar $Si_3N_4$ waveguides [52, 56]. In contrast, the fit $γ$'s of the hybrid waveguides with 1 and 2 layers of GO were ~11.7 $W^{-1}m^{-1}$ and ~27.9 $W^{-1}m^{-1}$, respectively, reflecting the significantly improved optical nonlinearity enabled by incorporating the highly nonlinear GO films. This is further verified by the more dramatic spectral broadening within the GO-coated

segment (i.e., window opening area) in comparison to the uncoated segments. The BWs of the simulated output spectra from the uncoated and hybrid waveguides with 1 and 2 layers of GO were ~91.7 nm, ~104.9 nm, and ~219.6 nm, respectively, in good agreement with the measured output spectra (i.e., ~90.9 nm, ~104.0 nm, and ~217.5 nm) in **Figure 6(a)**.

Based on the SCG modeling in **Eq. (3)** and the fit parameters obtained from **Figure 8**, we further analyze the influence of GO film length ($L_c$) and coating position ($L_0$) on the SCG performance. **Figures 9(a)** and **(b)** show the calculated BWs and BRs of the output spectra from the hybrid waveguides versus $L_c$ and peak power ($P_{peak}$) of input optical pulses, respectively. In each figure, (i) and (ii) show the results for the hybrid waveguides with 1 and 2 layers of GO, respectively. The coating position is fixed at $L_0 = 0.7$ mm – the same as those of the fabricated devices in **Sections 3** and **4**. The black points mark the results corresponding to the SCG measurements in **Figure 6(a)**, and the black triangles mark those corresponding to the maximum values of the -15-dB BWs or BRs. As can be seen, the BW increases with both $L_c$ and $P_{peak}$. For the hybrid device with 1 layer of GO, the maximum BW of ~262.2 nm and BR of ~3.6 are achieved at $L_c = 19.3$ mm and $P_{peak} = 1500$ W. Whereas for the hybrid device with 2 layers of GO, the maximum BW and BR are ~610.8 nm and ~8.4, respectively. This reflects that there is a large room for improvement in the SCG-induced spectral broadening by further increasing the GO film length and the input peak power.

**Figures 9(c)** and **(d)** show the calculated BWs and BRs of the output spectra from the hybrid waveguides versus coating position $L_0$ and $P_{peak}$, respectively. The film length is fixed at $L_c = 1.4$ mm, which is the same as those of the fabricated devices in **Sections 3** and **4**. For the hybrid device with 1 layer of GO, the maximum BW of ~126.2 nm and BR of ~1.7 are achieved at $L_0 = 0$ mm and $P_{peak} = 1500$ W. Whereas for the hybrid device with 2 layers of GO, the maximum 15-dB BW and BR are ~282.5 nm and BR of ~3.9, respectively. Both the BW and BR increase with $P_{peak}$ but decrease with $L_0$. The former shows a trend similar to that in **Figures 9(a)** and **(b)**, whereas the latter shows a trend opposite to their change with $L_c$. This is because the spectral broadening is mainly achieved within the GO coated region and the light power at the start of this region decreases with $L_0$.

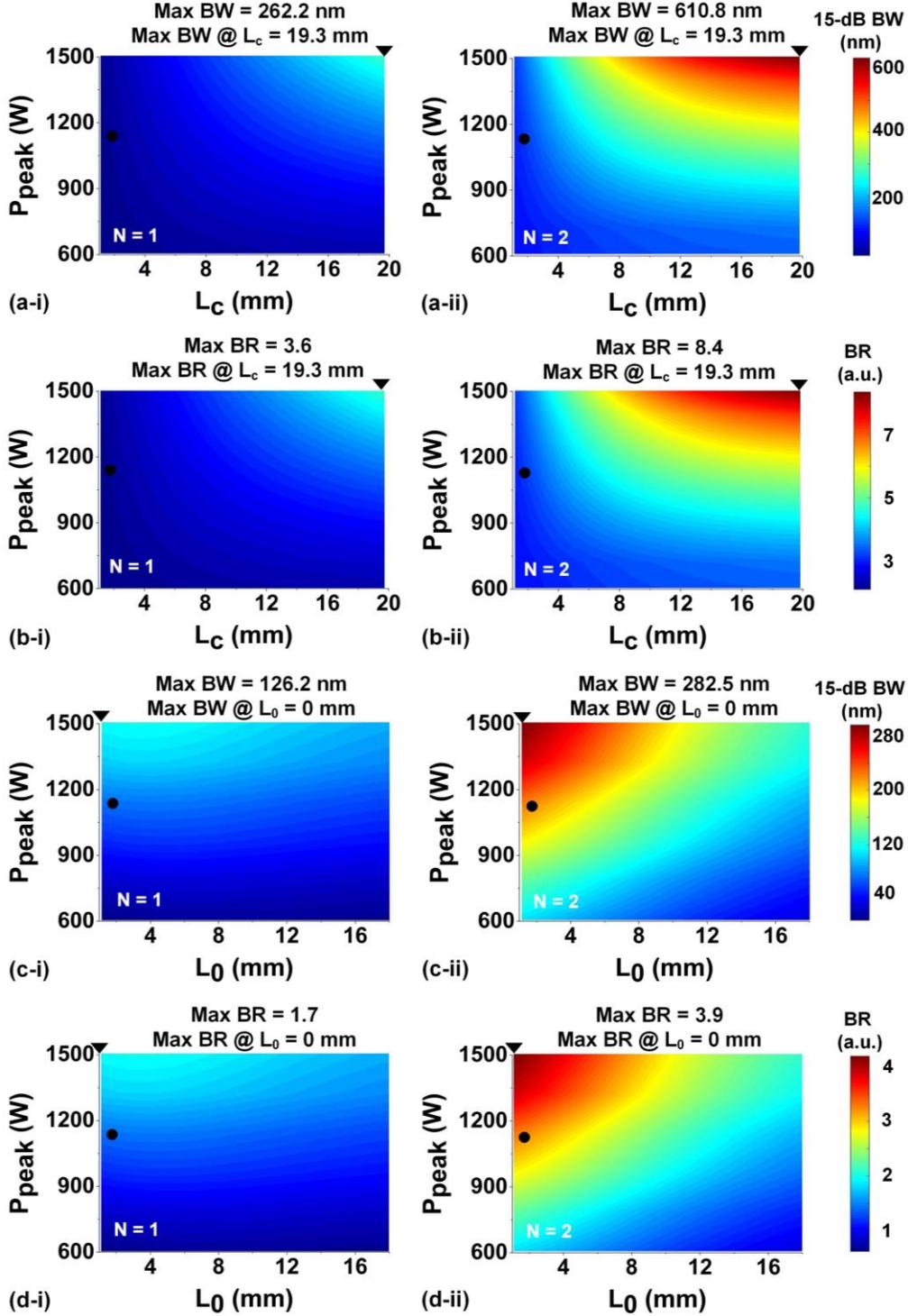

**Figure 9**. (a) Simulated 15-dB bandwidths (BWs) of the output spectra from the hybrid waveguides versus GO film length ($L_c$) and peak power of input optical pulses ($P_{peak}$). (b) Simulated broadening ratio (BRs) of the output spectra from the hybrid waveguides versus $L_c$ and $P_{peak}$. (c) Simulated 15-dB BWs of the output spectra from the hybrid waveguides versus GO coating position ($L_0$) and $P_{peak}$. (d) Simulated BRs of the output spectra from the hybrid waveguides versus $L_0$ and $P_{peak}$. In (a) − (d), (i) and (ii) show the results for the hybrid waveguides with 1 and 2 layers of GO ($N$ = 1, 2), respectively. The black points mark the results corresponding to the SCG experiments in Figure 6(a), and the black triangles mark the results corresponding to the maximum values of 15-dB BWs and BRs. All the devices have the same geometry ($W$ = 1.60 μm, $H$ = 0.72 μm) for the $Si_3N_4$ waveguides. In (a) and (b), $L_0$ = 0.7 mm. In (c) and (d), $L_c$ = 1.4 mm.
Internal

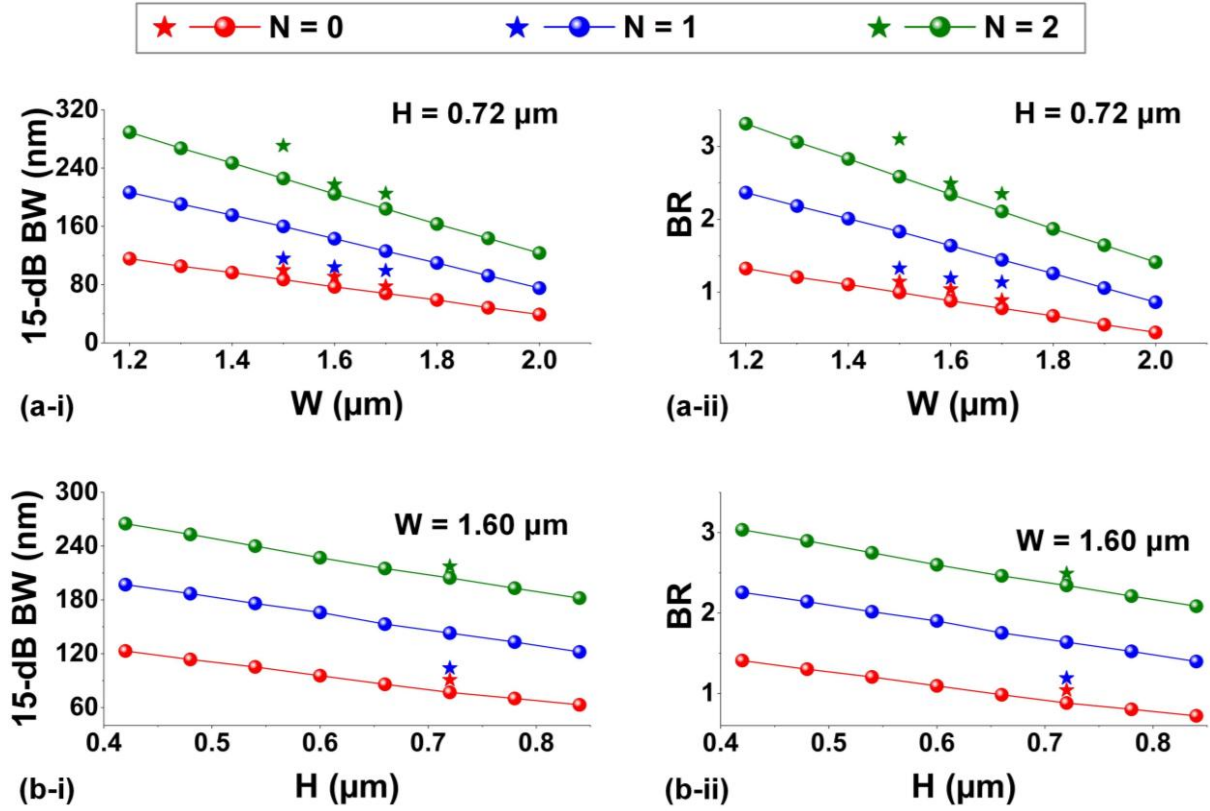

**Figure 10**. (a) Simulated 15-dB bandwidths (BWs) and broadening ratios (BRs) of the output spectra from the uncoated and hybrid waveguides with 1 and 2 layers of GO versus waveguide widths ($W$). (b) Simulated 15-dB BWs and BRs of the output spectra from the uncoated and hybrid waveguides with 1 and 2 layers of GO versus waveguide heights ($H$). In (a) and (b), (i) and (ii) show the results for the 15-dB BWs and the BRs, respectively. The stars mark the results corresponding to the SCG experiments in Figure 7. The peak power of input optical pulses is ~1100 W. The GO film length and coating position are $L_c$ = 1.4 mm and $L_0$ = 0.7 mm, respectively. In (a), the waveguide height is $H$ = 0.72 μm. In (b), the waveguide width is $W$ = 1.60 μm.

We also investigate the influence of the waveguide geometry ($W$, $H$) on the SCG performance. **Figure 10(a)** shows the simulated BWs and BRs of the output spectra from the uncoated and hybrid waveguides with 1 and 2 layers of GO versus different waveguide widths ($W$) at a fixed height $H$ = 0.72 μm. The stars are the SCG measurements in **Figure 7**. As can been seen, the BWs and BRs increase as $W$ decreases due to the significantly improved Kerr nonlinearity arising from the increased GO mode overlap. **Figure 10(b)** shows the corresponding results for different $H$ at a fixed width $W$ = 1.60 μm. Due to the increased GO mode overlap as $H$ decreases, the BWs and BRs decrease as $H$ increases, showing a trend similar to that in **Figure 10(a).** We note that the simulation results show some deviation from the experimental results, which is mainly induced by the differences between the experimentally generated input optical pulses and the perfect Gaussian input optical pulses assumed in our simulations.

In our experimental demonstration, the spectrally broadened output signals after propagation through the hybrid waveguides achieved a maximum 15-dB bandwidth of ~217.5 nm, which is comparable to those of the supercontinuum spectra in Refs. [16, 18, 38]. We note that supercontinuum spectra spanning over an octave in width are more versatile for many applications [2, 35, 41]. In this work, the experiments were mainly used as a proof-of-concept demonstration for comparing the SCG performance of the bare and hybrid waveguides. The bandwidths of our SCG can be significantly increased with increased peak power, such as the sources used in Refs. [32, 41, 84]. According to the simulation results in **Figures 9** and **10**, the supercontinuum spectra can also be broadened by optimizing the structural parameters of the GO-$Si_3N_4$ hybrid waveguides.

This is the most recent paper in our work on GO films integrated with waveguides [85-87]. We anticipate that it will be applicable for mid-IR applications [88, 89] and that it will be highly relevant for ultrahigh bandwidth nonlinear optical applications of microcombs for both classical [90-138] and quantum [139-150] applications.

## 6. Conclusion

In summary, we investigate SCG in $Si_3N_4$ waveguides integrated with 2D GO films and experimentally demonstrate significantly enhanced SCG compared to uncoated waveguides. We fabricate hybrid waveguides with precise control of the thickness, coating length and position of the GO films. SCG measurements using optical pulses with ultrahigh peak powers exceeding 1000 W are performed showing a significantly improved spectral broadening (up to 2.4 times in the -15-dB bandwidth) for the hybrid devices relative to the device without GO. Based on the experimental results, the influence of GO film thickness, coating length, coating position, and waveguide geometry on the SCG performance is theoretically analyzed, showing that further improvement can be achieved by optimizing the device structural parameters. These results confirm the effectiveness of on-chip integration of 2D GO films to enhance the SCG performance of photonic integrated devices for broadband nonlinear optical applications.